\documentclass[pra,showpacs,floatfix,onecolumn]{revtex4}
\usepackage{graphicx}
\usepackage{epstopdf}
\usepackage{amssymb}
\usepackage{amsmath}
\usepackage{xspace}
\usepackage{color}

\begin{document}

\title{Generation and protection of Maximally Entangled State between many
modes in an optical network with dissipation}
\author{Raul Coto}
\affiliation{Instituto de F\'{\i}sica, Pontificia Universidad Cat\'{o}lica de Chile,
Casilla 306, Santiago, Chile}
\author{Miguel Orszag}
\affiliation{Instituto de F\'{\i}sica, Pontificia Universidad Cat\'{o}lica de Chile,
Casilla 306, Santiago, Chile}
\author{Vitalie Eremeev}
\affiliation{Facultad de Ingenier\'ia, Universidad Diego Portales, Santiago, Chile}

\date{May 10, 2016}
\begin{abstract}
We present a three-cavity network model with two modes in each cavity and a
non-linear medium that generates a Kerr type interaction via both self-phase
and cross-phase modulation processes. We have two main goals. The first one
is to generate a multipartite Maximally Entangled State (MES), starting from
the ground state of the system. We address the problem both without and with
dissipation. Secondly, we want to protect the MES from decoherence. While
studying the MES, we analyze different bipartite and multipartite
entanglement measures. We also study the effect of an Avoided Level Crossing
(ALC) identified by the critical behavior of the entanglement measures, thus
showing that the quantum correlations act as a witness for such phenomena. Our
findings provide the quantum tools to perform the operation of generation
and protection of a maximally entangled state in a cavity QED environment.
\end{abstract}

\maketitle

\section{Introduction}

At the present time, quantum entanglement \cite{Horodecki} is still the most
common and efficient resource \cite{Vedral} for quantum information,
computation and communication tasks, thus it is highly desirable to operate
with maximally entangled states, which can be realized in a variety of
setups \cite{Horodecki, Zhao, Kastoryano}. Hence, quantum protocols used for
generation, protection and communication of the Maximally Entangled
State(MES) are in the continuous development and improvement, even
considering that during the last decade, alternative resources, such as
quantum discord, have been put forward and studied theoretically as well as
experimentally \cite{Datta, Piani1, Xu}.

Furthermore, all non-classical correlations can be activated into
distillable entanglement and recently it has been shown that one can
generate entanglement from classical correlations via local dissipation \cite%
{Piani2,Streltsov, Adesso, Lanyon, Orieux}.

Besides generation of entanglement, e.g. MES, another trascendental task in
quantum physics is the protection of this resource from the effects of
decoherence and dissipation, which naturally destroy partially (for short
times) and totally (for long times) the quantum correlations. Alternatively
to the protection by the common methods as isolation, error correction \cite%
{Nielsen, Zeng, Devoret, Reed}, decoherence-free subspace \cite{Lidar,
Mundarain}, etc., a kind of counteroffensive approach has been used lately,
known in literature as quantum bath engineering (QBE) or engineered
dissipation \cite{Poyatos, Verstraete}, i.e. a procedure which permits
driving an open quantum system to target states (correlated, coherent, etc.)
by engineering the mechanisms of dissipation and decoherence. In this
direction, during a relatively short period of time, many interesting
theoretical and experimental studies proved that this strategy works well
and can be very efficient for various physical systems \cite{Kastoryano,
Verstraete, Barreiro, Krauter, Murch, Kordas, Aron, Holland, Leghtas,
Mikhalychev, Mogilevtsev, Chaitanyaa, Azouit}. In this order of ideas, we
present here a new example, applying efficiently the principle of QBE for
the studied model, as will be explained in detail in the Section 3. In
general lines, our proposal of QBE is based on a kind of nonlinear
(two-mode) dissipation to the reservoir. Similar models \cite{Leghtas,
Mikhalychev, Mogilevtsev, Chaitanyaa, Azouit} have been used recently.

In the present paper we investigate a theoretical model of a cavity network
involving photonic Kerr non-linear effect, which provides the generation of
MES in such a setup. Particularly, by a non-adiabatic evolution of the
system from its ground state, possessing a very small fraction of
entanglement, the preparation of MES is possible, as shown via the fidelity
and negativity measures. We show that, the phenomenon of MES is conceivable
under the conditions of closed and open system, e.g. considering the losses
of the photons from the cavities. We also demonstrate how the MES could be
protected in the open system, using a particular two-mode coupling to the
environment, in such a way that it is possible to almost freeze the MES in
the case of a phase flip noisy channel. Moreover, phenomenon of avoided
level crossing (ALC) \cite{Heiss, Bhattacharya, Eleuch} appears in our
model, and we study how this effect is related to the quantum correlations
in the system, effect already discussed in the literature \cite{Reed,Gonzalez,
Karthik,Oh,Wang}. Therefore, the main purpose of this work is to advance
the field of quantum engineering, where original ideas for the production,
control and protection of photonic MES are suggested and developed in a
network of optical cavities under the approaches of closed and open system.

The remainder of this paper is structured in three sections.

In Sec. 2 we present a network model as a closed quantum system, defined in
general for N cavities and numerically analyzed for the case of three
cavities than could be, of course extended for more insight. We study the
preparation of the MES and witness the entanglement through the Negativity
and Fidelity. When the system evolves adiabatically, the phenomenon of
avoided crossing is identified by the critical behavior of the bipartite and
multipartite entanglement. These effects are illustrated and discussed.

In Section 3, the model of a cavity network is studied in the framework of
an  open quantum system, by considering two different damping mechanisms:
(i) arbitrary photons of mode $a$ and $b$ leave the cavity at rates $\gamma
^{a}$ and $\gamma ^{b}$, respectively; and (ii) the two-mode paired photons
abandon the cavity at rate $\gamma $. Both damping processes occur under the
condition of thermal environments at zero temperature. As a result, it is
shown that by the damping of photons correlated in pairs, the MES decays
slower in time, so concluding that such a bath engineering makes MES more
robust. Moreover, if the decoherence to such environment corresponds to a
phase flip noise channel, Eq.(\ref{phase_p}), then MES remains a steady
state, evidencing its maximal fidelity. The last Section is devoted to the
Conclusions.


\section{The cavity network model without losses}

We have an array of $N$ two-mode cavities and inside of each one there is a
non-linear medium which introduces a field-field interaction by Kerr
self-phase modulation \cite{Joshi} and Kerr cross-phase modulation process 
\cite{Harouni}. Furthermore, photons can hop between nearest neighbor
cavities. The total Hamiltonian (in units of $\hbar $) consists in three
parts: a free part $H_{0}$, a hopping term $H_{hop}$ and photon-photon
interaction term $H_{int}$.

\begin{eqnarray}  \label{hamiltonian}
H_0 &=& \sum_{i=1}^N \big [ \omega_i^a a_i^{\dagger}a_i + \omega_i^b
b_i^{\dagger}b_i \big ]  \label{H0} \\
H_{int} &=& \sum_{i=1}^N \big [ k_a (\hat{n}^a_i)^{2} + k_b (\hat{n}%
^b_i)^{2} + k_{int}\hat{n}_i^a \hat{n}_i^b \big ]  \label{Hf} \\
H_{hop} &=& \sum_{i=1}^{N-1}J_i \big [a_i^{\dagger} a_{i+1}e^{i\phi_a} +
b_i^{\dagger} b_{i+1}e^{i\phi_b} + h.c.\big ]  \label{Hhopp}
\end{eqnarray}

Notice that the interaction Hamiltonian (\ref{Hf}) \cite{Yamamoto,Aliaga},
indicating the contribution of the Kerr medium, involves only quadratic
elements. The first two terms correspond to the self-phase modulation
process, and usually appear as $(a^{\dagger })^{2}(a)^{2}$, but this product
of the creation and annihilation operators can be reordered using the
commutation relationship to get Eq.(\ref{Hf}), which yields constants and
linear terms which can be neglected since they commute with the Hamiltonian.
The third term is related to the Kerr cross-phase modulation, and introduces
an effective interaction between the two modes. In order to simplify the
problem, we can eliminate $H_{0}$ going to the interaction picture with the
unitary transformation $U=e^{-iH_{0}t}$, which leaves $H_{int}$ and $H_{hop}$
invariant under the conditions $\omega _{i}^{a}=\omega ^{a}$ and $\omega
_{i}^{b}=\omega ^{b}$. For the rest of the paper, our Hamiltonian will have
two parts only , $H_{hop}$ and $H_{int}$.


\subsection{Preparation of Maximally Entangled State (MES)}

The first objective of this study, is to prepare MES from an arbitrary
disentangled or partially entangled state by applying the above Hamiltonian
in the proposed photonic network. Such a MES have been extensively studied 
\cite{Horodecki}, and the general form of these states can be expressed by

\begin{equation}  \label{MES}
|MES\rangle = \sum_{\vec{n}}\lambda_{\vec{n}}|n_1,\dots,n_N\rangle_a\otimes
|n_1,\dots,n_N\rangle_b,
\end{equation}

where $|n_1,n_2,\dots,n_N\rangle_{a(b)}$ represents the state with $n_i$
photons in cavity $i$ for the mode $a$($b$), while $\vec{n}$ runs over all
possible combinations of $\lbrace n_1,\dots,n_N\rbrace$. It is important to
notice that there are the same number of photons for each mode in each
cavity. Next, we apply the Hopping Hamiltonian to this state:

\begin{eqnarray}
H_{hop}|MES\rangle  &=&\sum_{i=1}^{N-1}\sum_{\vec{n}}J_{i}\lambda _{\vec{n}%
}[e^{i\phi _{a}}\sqrt{n_{i}+1}\sqrt{n_{i+1}}|n_{1},\dots
,n_{i}+1,n_{i+1}-1,\dots ,n_{N}\rangle _{a}\otimes |n_{1},\dots
,n_{N}\rangle _{b}  \notag  \label{hopT1} \\
&+&e^{-i\phi _{a}}\sqrt{n_{i}}\sqrt{n_{i+1}+1}|n_{1},\dots
,n_{i}-1,n_{i+1}+1,\dots ,n_{N}\rangle _{a}\otimes |n_{1},\dots
,n_{N}\rangle _{b}  \notag \\
&+&e^{i\phi _{b}}\sqrt{n_{i}+1}\sqrt{n_{i+1}}|n_{1},\dots ,n_{N}\rangle
_{a}\otimes |n_{1},\dots ,n_{i}+1,n_{i+1}-1,\dots ,n_{N}\rangle _{b}  \notag
\\
&+&e^{-i\phi _{b}}\sqrt{n_{i}}\sqrt{n_{i+1}+1}|n_{1},\dots ,n_{N}\rangle
_{a}\otimes |n_{1},\dots ,n_{i}-1,n_{i+1}+1,\dots ,n_{N}\rangle _{b}].
\end{eqnarray}

Now, by replacing the indices $(n_i +1)\rightarrow n_i$ and $%
(n_{i+1}-1)\rightarrow n_{i+1}$, in the third term, and for the fourth term, 
$(n_i -1)\rightarrow n_i$ and $(n_{i+1}+1)\rightarrow n_{i+1}$, we readily
get,

\begin{eqnarray}
H_{hop}|MES\rangle  &=&\sum_{i=1}^{N-1}J_{i}\sum_{\vec{n}}[\lambda _{\vec{n}%
}e^{i\phi _{a}}\sqrt{n_{i}+1}\sqrt{n_{i+1}}|n_{1},\dots
,n_{i}+1,n_{i+1}-1,\dots ,n_{N}\rangle _{a}\otimes |n_{1},\dots
,n_{N}\rangle _{b}  \notag  \label{hopT2} \\
&+&\lambda _{\vec{n}}e^{-i\phi _{a}}\sqrt{n_{i}}\sqrt{n_{i+1}+1}|n_{1},\dots
,n_{i}-1,n_{i+1}+1,\dots ,n_{N}\rangle _{a}\otimes |n_{1},\dots
,n_{N}\rangle _{b}  \notag \\
&+&\lambda _{\{n_{i}-1,n_{i+1}+1\}}e^{i\phi _{b}}\sqrt{n_{i}}\sqrt{n_{i+1}+1}%
|n_{1},\dots ,n_{i}-1,n_{i+1}+1,\dots ,n_{N}\rangle _{a}\otimes |n_{1},\dots
,n_{N}\rangle _{b}  \notag \\
&+&\lambda _{\{n_{i}+1,n_{i+1}-1\}}e^{-i\phi _{b}}\sqrt{n_{i}+1}\sqrt{n_{i+1}%
}|n_{1},\dots ,n_{i}+1,n_{i+1}-1,\dots ,n_{N}\rangle _{a}\otimes
|n_{1},\dots ,n_{N}\rangle _{b}].
\end{eqnarray}

We can regroup Eq.(\ref{hopT2}) and take $J=J_{i\text{ }}$ such that

\begin{eqnarray}
H_{hop}|MES\rangle  &=&J\sum_{i=1}^{N-1}\sum_{\vec{n}}[(\lambda _{\vec{n}%
}e^{i\phi _{a}}+\lambda _{\{n_{i}+1,n_{i+1}-1\}}e^{-i\phi _{b}})  \notag
\label{hopT3} \\
&\times &\sqrt{n_{i}+1}\sqrt{n_{i+1}}|n_{1},\dots ,n_{i}+1,n_{i+1}-1,\dots
,n_{N}\rangle _{a}\otimes |n_{1},\dots ,n_{N}\rangle _{b}  \notag \\
&+&(\lambda _{\vec{n}}e^{-i\phi _{a}}+\lambda
_{\{n_{i}-1,n_{i+1}+1\}}e^{i\phi _{b}})  \notag \\
&\times &\sqrt{n_{i}}\sqrt{n_{i+1}+1}|n_{1},\dots ,n_{i}-1,n_{i+1}+1,\dots
,n_{N}\rangle _{a}\otimes |n_{1},\dots ,n_{N}\rangle _{b}].
\end{eqnarray}

One can see that it is possible to have zero hopping energy, by generating
vanishing prefactors in Eq.(\ref{hopT3}). Then, using this idea we get the
following condition for the first prefactor,

\begin{equation}  \label{cond1}
\lambda_{\vec{n}} e^{i(\phi_a + \phi_b+\pi)} = \lambda_{\lbrace
n_i+1,n_{i+1}-1\rbrace}.
\end{equation}

The previous condition indicates that a single hopping is equivalent to
introducing a phase $e^{i(\phi _{a}+\phi _{b}+\pi )}$.  Since the system is
periodic, we can repeat this process $N$ times to get,

\begin{equation}  \label{cond2}
\lambda_{\vec{n}} e^{iN(\phi_a + \phi_b +\pi)} = \lambda_{\vec{n}},
\end{equation}

and with the above result, the condition for having vanishing hopping energy
is,

\begin{equation}
(\phi _{a}+\phi _{b})=\frac{(2m-N)\pi }{N},  \label{cond3}
\end{equation}%
with $m$ an integer number. The same condition for the phases applies to the
other term in Eq.(\ref{hopT3}). This result is very important, summarizing
the first step towards the generation of a MES. The second step, is to
generate \ a zero interaction energy, then we apply the $H_{int}$ to the MES
state,

\begin{equation}  \label{HfT1}
H_{int}|MES\rangle = \sum_{\vec{n}}\sum_{i=1}^N \lambda_{\vec{n}}(k_a +k_b +
k_{int})n_i^2|n_1,\dots,n_N\rangle_a\otimes |n_1,\dots,n_N\rangle_b.
\end{equation}

It is worth noticing that the terms inside the parenthesis do not depend on $%
n_{i}$ because of the form of the MES state, where there is the same number
of photons in each mode. Following the approach in \cite{Aliaga}, the
authors showed that $k_{int}=2\sqrt{k_{a}k_{b}}$. The factor $k_{a(b)}$ is
proportional to the real part of the third-order susceptibility, $\chi
^{(3)} $, which can be negative \cite{Landauer, Sipe}. Thus, setting $%
k_{a}=k_{b}=-k$, we get zero eigenvalue for both the Interacion and full
Hamiltonian.

Finally, replacing condition (\ref{cond1}) in Eq.(\ref{MES}) we get the MES
definition,

\begin{equation}  
|MES\rangle = \sum_{\vec{n}}\frac{1}{\sqrt{d}}e^{i\frac{2\pi m}{N}p_{\lbrace
n_1,\dots n_N \rbrace}}|n_1,\dots,n_N\rangle_a\otimes
|n_1,\dots,n_N\rangle_b,
\end{equation}

where $p_{\lbrace \dots ,n_j+1, n_{j+1}-1,\dots \rbrace}=p_{\lbrace
\dots,n_j, n_{j+1},\dots \rbrace}+1$ and $d$ is the dimension of the Hilbert
space \cite{Reyes}.

One can reach the MES dynamically by varying the phase non adiabatically, as
depicted in Fig.1 by the red triangles, where we set $\phi_a=\phi_b=\phi$. When $m=3$ and $\phi=\pi/2$ the MES state that we reached is:

\begin{eqnarray}
\vert MES\rangle &=& \frac{1}{\sqrt{6}}(\vert 011\rangle\otimes \vert 011\rangle + \vert 101\rangle\otimes \vert 101\rangle +\vert 110\rangle\otimes \vert 110\rangle \nonumber \\&+&  \vert 200\rangle\otimes \vert 200\rangle + \vert 020\rangle\otimes \vert 020\rangle + \vert 002\rangle\otimes \vert 002\rangle).
\label{MES2}
\end{eqnarray}

 On the other hand, if the phase varies adiabatically, then the system follows its initial eigenstate, patterned by the blue squares. The effects of the avoided level crossing will be
discussed in the subsection C.


\begin{figure}[ht]
\centering
\includegraphics[width=0.5 \textwidth]{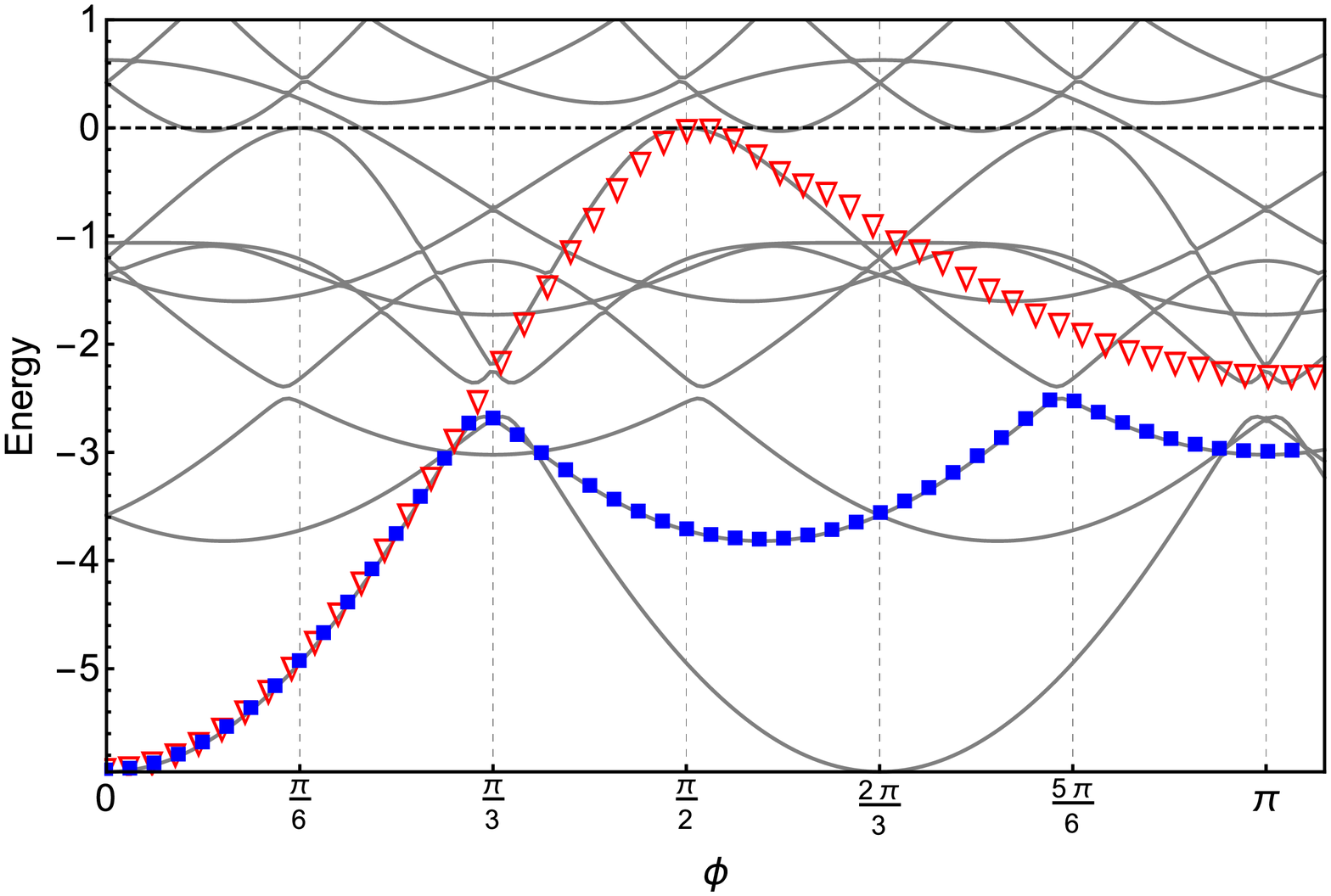}
\caption{Energy levels as function of the phase. Preparation of MES by non
adiabatic variation of phase (red triangles); ground state evolution in the
adiabatic passage (blue squares). Here $k_a=k_b=J$.}
\label{fig1}
\end{figure}


\subsection{MES: bipartite or multipartite correlations?}


We consider our present system composed of six qutrits (states with 0,1 and
2 photons) defined by three cavities and two modes per cavity. The quantum
correlations in a system of any dimension, i.e. qudits, are usually measured
by the Negativity \cite{negativity,negativity2}. In our particular case, in
order to quantify the amount of bipartite (two-body) quantum correlation, we
trace over four of the qutrits and use Negativity to calculate the
correlations between the residuary qutrits.

In general, for subsystems $X$ and $Y$ and an associated density matrix $%
\rho _{XY}$, the Negativity is defined as

\begin{equation}  \label{negat}
\mathcal{N}(\rho_{XY})=\sum_i \frac{\vert \lambda_i \vert - \lambda_i}{2},
\end{equation}
where $\lambda_i$ are the eigenvalues of the partial transpose of the
density matrix, $\rho^{T_{X(Y)}}$, with respect to one of the subsystems. It
essentially measures the degree to which $\rho^{T_{X(Y)}}$ fails to be
positive, and therefore it can be considered as a quantitative version of
Peres's criterion for separability \cite{peres}.
\begin{figure}[ht]
\par
\centering
\includegraphics[width=0.45 \textwidth]{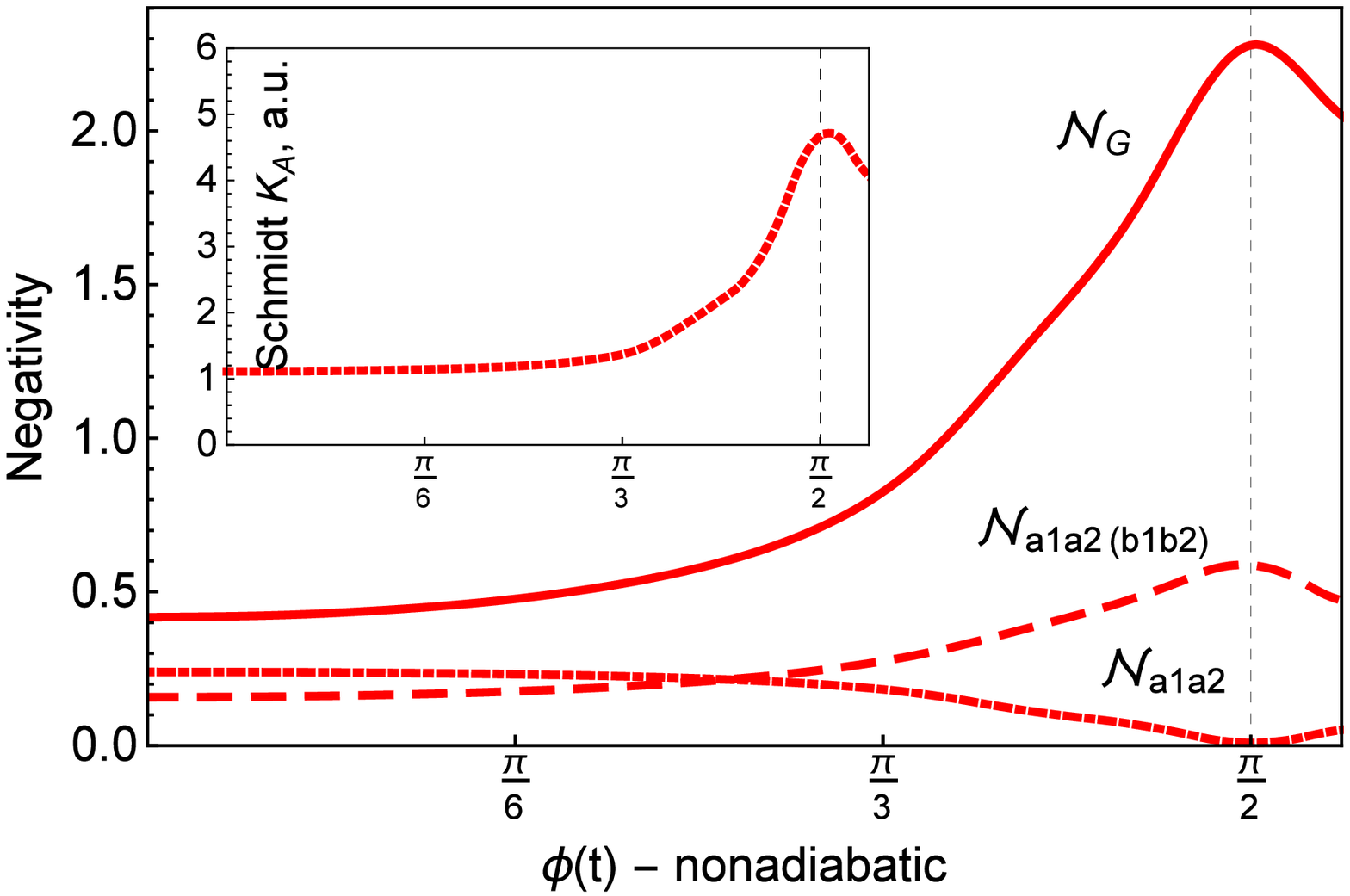}
\includegraphics[width=0.45 \textwidth]{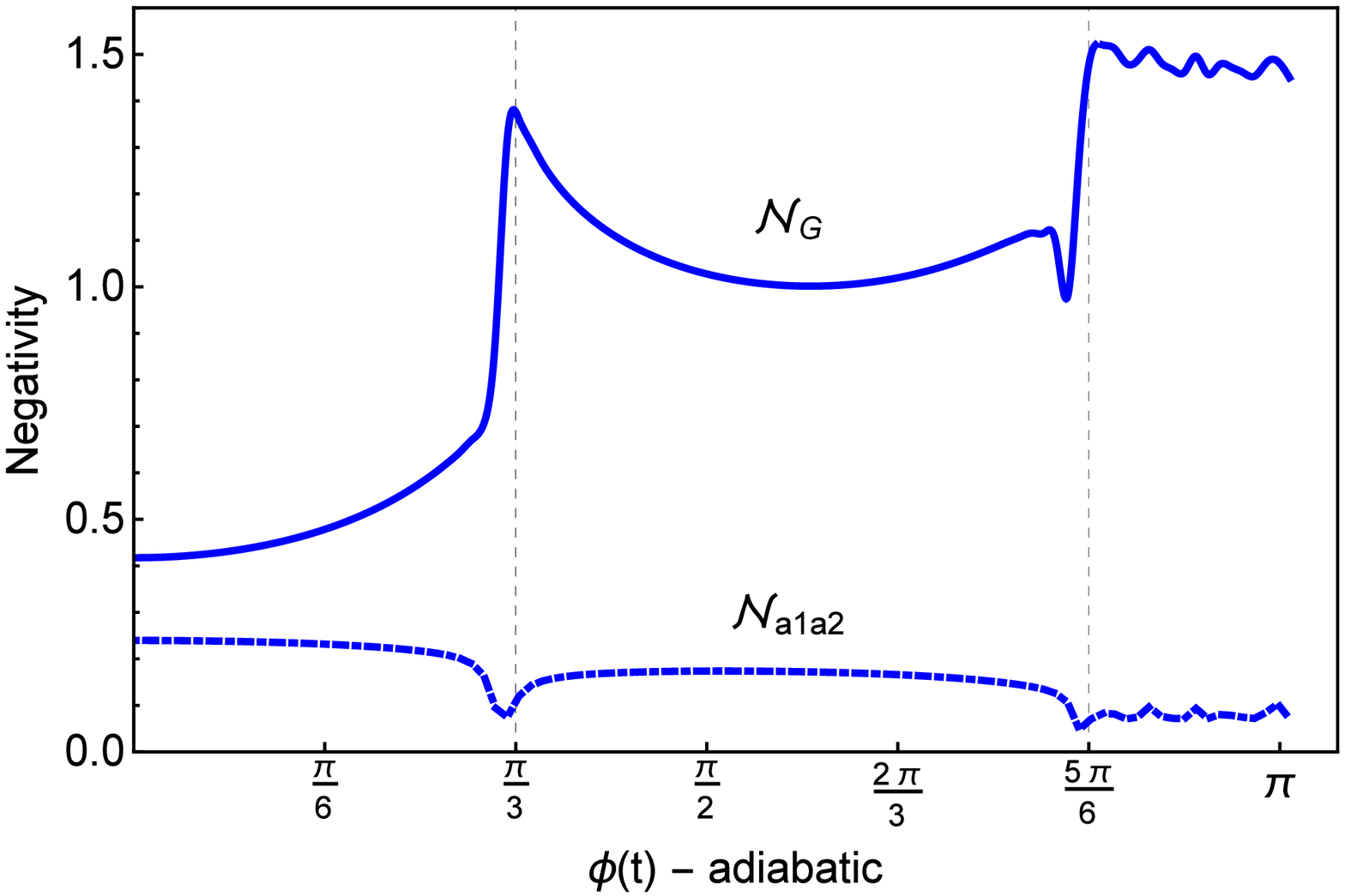}
\caption{(left) Global ($\mathcal{N}_{G}$) and partial ($\mathcal{N}%
_{a1a2(b1b2)}$) multipartite Negativity reach a maximum when approaching the
MES phase ($\protect\pi /2$), while bipartite correlations ($\mathcal{N}%
_{a1a2}$) vanish. Inset: evolution of $\mathcal{N}_{G}$ and Schmidt number
agrees. (right) Two avoided level crossing are witnessed by an abrupt
increment of $\mathcal{N}_{G}$, while decreasing bipartite correlations, $%
\mathcal{N}_{a1a2}$. The parameters are as in Fig.1.}
\label{fig2} 
\end{figure}

\begin{figure}[ht]
\centering
\includegraphics[width=0.5 \textwidth]{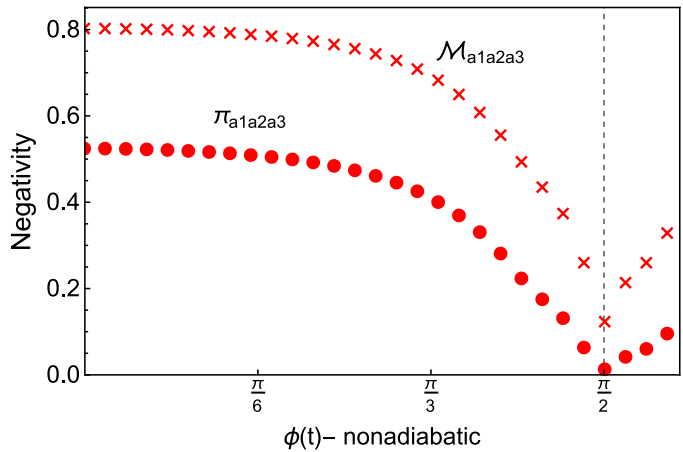}
\caption{Evolution of Negativity measuring the tripartite correlations using
Eq.(\protect\ref{NegPi}) (circles) and Eq.(\protect\ref{NegM}) (crosses) for
the nonadiabatical passage for the same parameters as in Fig.2.}
\label{fig3a}
\end{figure}

After preparing the target MES, it is important to evaluate the degree of
entanglement of this state. The MES defined by Eq.(\ref{MES2}) is mostly a
global (multipartite) entangled state between two different modes localized
in three cavities. This means that all possible bipartite correlations, say
mode $a$ and $b$ of the same cavity or different cavities have no
entanglement (zero Negativity). Negativity is only a sufficient condition
for entanglement, however, by tracing out over four of the qutrits, ending
with only two qutrits, one can realize that the final state has no quantum
correlation between its parts. Even more, the MES has no correlation between
subsystems of the same mode, not bipartite correlation ($\mathcal{N}_{a1a2}$%
), as shown by dot-dashed curve in Fig. \ref{fig2}a, nor tripartite
correlation ($\mathcal{\pi}_{a1a2a3}$) plotted in Fig. \ref{fig3a} \ and
shown by bullets. Nevertheless, we cannot say that the MES state belongs to
the $GHZ$ class of multipartite entangled states. We recall that for the GHZ
state, when eliminating one of the qudits, all correlations are destroyed,
ending with a state proportional to the identity matrix. In our case, if we
trace out over only one cavity, for example cavity three, we still find
correlations between the remaining modes, evidenced when calculating $%
\mathcal{N}_{a1a2(b1b2)}$ similar to Eq.(\ref{NG}) and represented with a
dashed curve in Fig. \ref{fig2}a.

Because of the high dimension of the system, to calculate any global
multipartite measure of correlations, considering both modes is a difficult
task.

For example, in order to measure tripartite correlations for the same mode,
we used two different definitions found in literature, as in \cite{Cheng} 
\begin{equation}
\pi _{a1a2a3}=\frac{1}{3}\sum_{i=1}^{3}\pi _{ai},  \label{NegPi}
\end{equation}

with $\pi_{a1}=\mathcal{N}_{a1(a2a3)}^2-\mathcal{N}_{a1a2}^2-\mathcal{N}%
_{a1a3}^2$; and as in \cite{Sabin}

\begin{equation}  \label{NegM}
\mathcal{M}_{a1a2a3}=\sqrt[3]{\mathcal{N}_{a1(a2a3)}\mathcal{N}_{a2(a1a3)}%
\mathcal{N}_{a3(a1a2)}}.
\end{equation}
In the next section (see Fig. \ref{fig3b}), we compare both measures
during the preparation of the MES state.

For a more general situation, in order to investigate the multipartite
correlations, we consider an approximate calculation, which can be carried
out in a relatively simple way, by considering the symmetry properties in
our model. For example, we may assume that for each mode one has an averaged
state for the three cavities which are identical, so instead of considering
three qutrits, there will be one qudit of dimension $d=3^{3}$. Applying the
same idea for the other mode, we end up with two qudits, for which may
calculate the standard negativity measure as in Eq.(\ref{negat}), as follows 
\begin{equation}
\mathcal{N}_{G}=\mathcal{N}_{a1a2a3(b1b2b3)},  \label{NG}
\end{equation}%
where the partial transpose is taken on mode $b$ in the three cavities.
Using this approach, we get a value that gives us a qualitative estimation
of the the global correlation between the two modes, considering that for
the MES, the bipartite and tripartite correlations vanish.

\subsection{Negativity witnessing MES and avoided level crossing}

In this subsection we analyze the quantum correlations measured by the
Negativity, and for verification the Schmidt number was computed as well. As
was explained above, to engineer the MES during the dynamics of the system
one should satisfy two conditions: (i) $k_{a}+k_{b}+k_{int}=0$ resulting
from Eq.(\ref{HfT1}), and (ii) managing the total phase given by Eq.(\ref%
{cond3}). The first condition can be achieved as in references \cite{Aliaga,
Landauer, Sipe}, and in our calculations one just fixes $%
k_{a}=k_{b}=-k_{int}/2$. In order to fulfill dynamically the second
condition, one could vary the phase both adiabatically and non-adiabatically
to meet the target value. 

Let`s start by managing the non adiabatic variation of the phase, so
initiating in the ground state, the system follows the track as represented
in Fig. \ref{fig1} by the red triangles. We see that for the phase equal to $%
\pi /2$ the system evolves to the state with zero energy, which could be the
MES. In order to check this, we compute the value of entanglement as shown
in Fig. \ref{fig2}a. In the main plot we show three curves, where the
continuous line represents the global (multipartite) entanglement between
the modes $a$ and $b$ in three cavities, the dashed line depicts the
entanglement between the modes $a$ and $b$ in two cavities ($\mathcal{N}%
_{a1a2(b1b2)}$), i.e. the partial multipartite correlation, and finally the
dot-dashed curve evidences the bipartite entanglement of the mode $a$ for
two different cavities. The results are quite clear. We find that the
maximal value of the multipartite entanglement measured by the Negativity as
well the Schmidt number (inset of Fig. \ref{fig2}a) occur in the region where
the phase is $\pi /2$, so witnessing the MES as given in Eq.(\ref{MES2}). On
the other hand, the minimal value of the bipartite entanglement indicates
that the MES has genuinely multipartite correlations, distributed between
the modes $a$ and $b $, additionally evidenced by the partially multipartite
entanglement as shows the dashed curve. We also computed the Negativity, not
shown here, for the bipartite states of modes $a$ and $b$ in the same or
different cavities, but the results show that the entanglement is zero at
all times. For a deeper insight on the correlations, we computed the
tripartite correlations for the mode a and represent numerically the
comparison between the two different methods as shown in Fig. \ref{fig3a},
where we see that the tripartite entanglement computed by Eq.(\ref{NegPi})
becomes zero in $\pi /2$ indicating once more that the MES has multipartite
correlations between the modes, hence one can conclude that the method
proposed in \cite{Cheng} is more appropriate, at least for our system.

Next, we continue by studying the adiabatic variation of the phase, as is
represented in the Fig. \ref{fig1} by the blue squares path. We readily
notice that the system follows its ground state during all the evolution and
never reach a MES. However, we observe another effect occurring during the
dynamics of the eigenstate, the so-called avoided level crossing (ALC). The
ALC is a region where two or more energy levels of the system are close
enough such that the dynamics might follow one path with almost no change in
the wave function or jump to a fairly different state (that occurs in the
case of non adiabatic variation of phase). The selection of one path or the
other can be tuned by changing one parameter, which increases (decreases)
the speed at which the wave function undergoes to the avoided crossing
leading to the non adiabatic (adiabatic) passage, see Fig. \ref{fig1}. It is
noted that when approaching the ALC, the evolution of the state turns out to
be very complex leading some times to chaos \cite{Heiss, Haake, Karthik} or
quantum phase transitions \cite{Sachdev}. The signature of an ALC is not
always easy to establish by just looking at the time evolution of the wave
function. Quantum correlations (QC) play an important role in detecting the
criticality \cite{Wu,Campbell, Werlang}. It has been shown that for
different systems, the QC change considerably in this region, and these may
reach an extremal value \cite{Gonzalez, Karthik, Oh,Wang, Campbell}. For
multiparticle systems, the real witness is the multipartite rather than
bipartite correlation. Furthermore, it has been pointed out for a spin chain
that at an ALC, the multipartite correlation increases while the bipartite
correlation decreases \cite{Karthik}. This behavior indicates that in this
region a collective effect shows up (global correlation), rather than a
nearest-neighbor dynamics (two-body interaction).

In order to establish with more certainty the effect of ALC in our model, we
analyzed the dynamics of the entanglement measured by Negativity. In Fig. %
\ref{fig1} we find that ALC seems to occur in the regions of $\phi \approx
\pi /3$ and $\phi \approx 5\pi /6$. The evolution of the Negativity as seen
in Fig. \ref{fig2}b, shows clearly the position of the ALC and is perfectly
witnessed by an abrupt change in the dynamics, evidencing maximal values for
multipartite entanglement and minimal values for bipartite entanglement. It
should be mentioned that the Schmidt number has very similar behavior. Our
results are in good agreement to a similar effect observed in \cite{Gonzalez}
and confirm the more general conclusions given in \cite{Karthik}, hence
shedding more light on the importance of QC in witnessing phenomena like
ALC, etc.

\begin{figure}[ht]
\centering
\includegraphics[width=0.5 \textwidth]{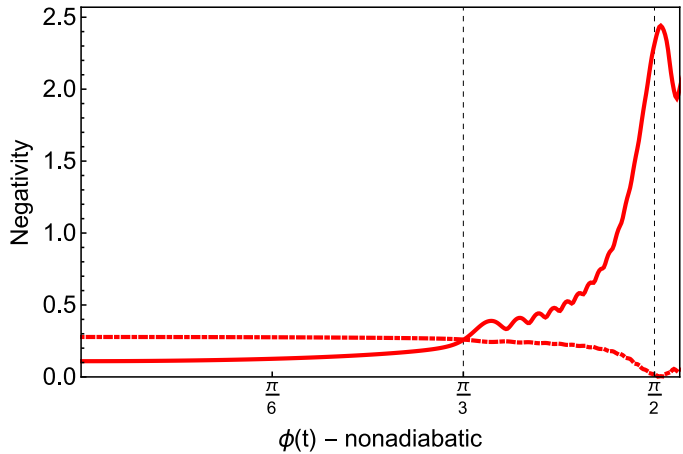}
\caption{Global entanglement (solid line) and bipartite entanglement
(dot-dashed line) similar as in Fig. 2a, except that $k_a=k_b=J/4$.}
\label{fig3b}
\end{figure}

To conclude this section, we find and emphasize here the importance of the
adiabatic and non adiabatic passages in our model. Even if the adiabatic
path that does not lead to a considerable change in the energy, there is a
sudden transition in the evolution of interspecies multipartite entanglement
(modes $a$ and $b$ in whole network) and inter-site bipartite entanglement
(same mode in two different cavities), effect shown in Fig. \ref{fig2}b,
which is similar to the one described by Karthik \textit{et al.} \cite%
{Karthik}. In the nonadiabiatic passage leading to the MES, one would also
expect to observe a particular behavior of the Negativity which changes
spontaneously its value in the region of ALC or \textit{exceptional point},
as named sometimes in literature \cite{Heiss, Bhattacharya, Eleuch}. This
effect is not observed well in Fig. \ref{fig2}a but is better visible for
the smaller gaps between the levels at the ALC, which could be managed, e.g.
by decreasing the rate $k/J$ in Eq.(\ref{Hf}) (see Fig. \ref{fig3b}, where
a more abrupt increase of the global entanglement is shown from the ALC
region on, as the path reaches the MES in its dynamics).


\section{Damping Effects in the photonic network}


\subsection{Preparation of a MES with losses}

In Fig. \ref{fig1} we showed that the MES (\ref{MES2}) can be prepared from
the ground state by changing the phase, which as a function of time can be
explicitly written as $\phi =\alpha t$. Notice that $\alpha $ will give the
speed of the system. When $\alpha $ is small enough, the system will follow
the adiabatic path, otherwise it will follow the non adiabatic path.
However, not all the non adiabatic paths will lead to the MES, actually,
there is a small range of values of $\alpha $ ($\alpha _{opt}$) that
optimize the MES preparation, reaching a high Fidelity ($\mathcal{F}=tr[\rho
|MES\rangle \langle MES|]$) \cite{Schumacher}.  We found that for $k=J/16$
and $\alpha_{opt}=3*10^{-4} J$ the Fidelity is $\mathcal{F}\approx 0.98$.
The nonlinear interaction $k$ is related to the height of the gap at the
avoided crossing. If this gap is considerably diminished such that even for
a slow tuning of the phase, \textit{i.e.} small $\alpha$, the system still
goes through the non adiabatic path, then the probability of reaching the
MES is very high. That is why we decreased $k$ down to $J/16$.
Unfortunately, $\alpha_{opt}$ depends on different parameters, as $k$, $J$
and $\gamma$ if losses were considered. Then, by decreasing $k$ we have to
decrease $\alpha$ too, in order to find $\alpha_{opt}$. For the case where
the interaction with the environment is not negligible, this mechanism is
not effective, since we need to vary the phase  very slowly, and eventually
losses will lead the state to a different path before becoming a MES. In the
presence of losses, it would be convenient for the system to quickly reach
the phase $\pi/2$. Another approach is to increase $k$ with respect to $J$.
In this case, as we decrease the hopping strength the inter-site correlation
becomes smaller, while the inter-species interaction gets stronger.  Under
this condition the energy spectrum shrinks down, and the ground state (from
which we start in the dynamics leading the MES) becomes very close to the
zero energy level, thus closer to the MES. This means that for large $\alpha$%
, which implies going faster to the MES, one still gets a good Fidelity ($%
\mathcal{F}\approx 0.9$) while at the same time losses are reduced. Another
way to understand this, is by comparing the hopping Eq.(\ref{Hhopp}) and the
interaction Eq.(\ref{Hf}) Hamiltonians. When $k$ gets bigger the interaction
term dominates over the hopping.  Because of this, the dependence with $%
\alpha$, which appears in the condition $(\ref{hopT3})$, is not that
relevant, since the condition ($\ref{HfT1}$) dominates and it is independent
of $\alpha$.

\begin{figure}[t]
\centering
\includegraphics[width=0.5 \textwidth]{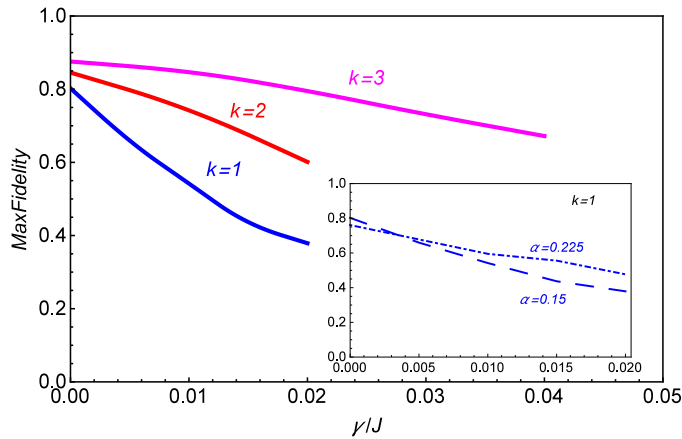}
\caption{The losses rate $\protect\gamma$ deteriorates the preparation of
the MES, but this can be overcome by increasing the nonlinear interaction $k$%
. Inset: Another approach to deal with losses is to increase the speed $%
\protect\alpha$, which for higher losses increases the Fidelity.}
\label{fig6}
\end{figure}

In order to simulate losses, we assume that each mode interacts with its own
reservoir at zero temperature, making the photons jump out of the cavities.
We use the master equation (ME) approach \cite{Orszag}. The equation of
motion for the density operator is given by,

\begin{equation}
\dot{\rho} = -i[H,\rho] +\mathcal{L}_a(\rho) +\mathcal{L}_b(\rho) ,
\end{equation}
where 
\begin{equation}  \label{ME}
\mathcal{L}_{\Lambda}=\sum_{k=1}^3 \frac{\gamma_k^{\Lambda}}{2}(2{\Lambda}%
_k\rho {\Lambda}_k^\dagger -\lbrace {\Lambda}_k^\dagger {\Lambda}_k,\rho
\rbrace) ,
\end{equation}

with the operator $\Lambda =\{a,b\}$, $k=\{1,2,3\}$ representing different
cavities, and $\gamma ^{a}$($\gamma ^{b}$) the decay rate for mode $a$($b$). 

In Fig. \ref{fig6} we show the variation of the maximum of the Fidelity with
respect to the MES as a function of the losses rate ($\gamma $). As one
would expect, as we increase the loss rate, the maximum Fidelity decreases.
However, in the main plot we see that by increasing the nonlinear
interaction $k$, holding $\gamma $ fixed, we manage to generate the MES with
higher Fidelity. Notice that each curve is plotted for its corresponding $%
\alpha_{opt}$ at $\gamma=0$. As we said above, $\alpha_{opt}$ is a function
of $\gamma$, but we will discuss this case later on. We found numerically
that there is a critical rate ($\gamma _{c}$), above which the Fidelity only
decreases, \textit{i.e.}, there is no global maximum for the Fidelity at $%
\pi /2$, evidencing that the system will not get closer to the MES with the
given set of parameters. For example, for $\alpha =0.15J$ and $k=J$ we found
that $\gamma _{c}\approx 0.02J$. This means that if in the system losses
cannot be controlled to be below this threshold, the only way to get the MES
is by tuning the other parameters.

Let's now discuss the dependence of $\alpha_{opt}$ with $\gamma$. By holding 
$k$ constant, as losses increase one can increase the speed $\alpha$ to
obtain higher Fidelity. As we explain above, the reason why this happens is
that the system will get faster to the phase $\pi/2$, which means going to
the MES before losses take it through a different path. However, this
behavior is not well evidenced for large $k$ ($k\geq 3 $), since the system
becomes insensitive to the variation of $\alpha$. Then, the consideration
of the dependence $\alpha_{opt}(\gamma)$ is  better observed for smaller $k
$. In the Inset of Fig. \ref{fig6}, we compare the maximum of the Fidelity
as a function of $\gamma$ for $\alpha=0.15$ (bottom curve of the main plot)
and $\alpha =0.225$. We chose for this plot $k=J$. We see that because of $%
\alpha=0.15$ (dashed line) is the optimal value at $\gamma=0$, this curve
starts above $\alpha=0.225$ (dot-dashed line). Nevertheless, when increasing
the losses both curves intercept and beyond that $\alpha=0.225$ becomes the
new optimal value. This method works as a mechanism for dealing with losses.

\subsection{Robustness against noise}

In this subsection, we are interested in preserving the MES (\ref{MES2}). In
order to counteract the effects of the environment, at least for a short
time of the evolution, a quantum error correction protocol \cite{Nielsen}
could be implemented. However, this is not a feasible solution, since it
will require the encoding of a very large state \cite{Zeng}, which is not
practical. A Decoherence-Free Subspace (DFS) \cite{Lidar, Mundarain} does
not lead to any significant result either, since a common reservoir for the
whole system, as well as an enlargement of the system would be needed. An
interesting approach to protect the MES against decoherence could rely on
reservoir engineering. Let's assume that each cavity interacts with its own
reservoir, but the two modes inside a cavity will interact in a certain way
such that if one experiences a jump, the other mode will follow the first
one, similar as in \cite{Chaitanyaa}. However, the non-linear losses not appear naturally in the standard optical cavities/resonators and should be artificially stimulated in order to compete with the single photon damping, usual for such devices. Experimentally, such non-linear damping is not an easy task, nevertheless an intuitive experiment could be considered were the damping occurs in an absorbing medium resonant with two-photon transitions, e.g. for our model. More complex experiment was proposed very recently \cite{Leghtas}, were the authors realize two-photon dissipation at the rate of the same order of magnitude as the single-photon decay rate (i.e. $\gamma$ is approximately similar in Eqs. (\ref{ME}) and (\ref{ME2})). Even more, the authors conclude: ``The ratio between these two rates can be further improved within the present technology by using an oscillator with a higher quality factor and increasing the oscillator's nonlinear coupling to the bath" (page 857, last paragraph).

\begin{figure}[ht]
\centering
\includegraphics[width=0.5 \textwidth]{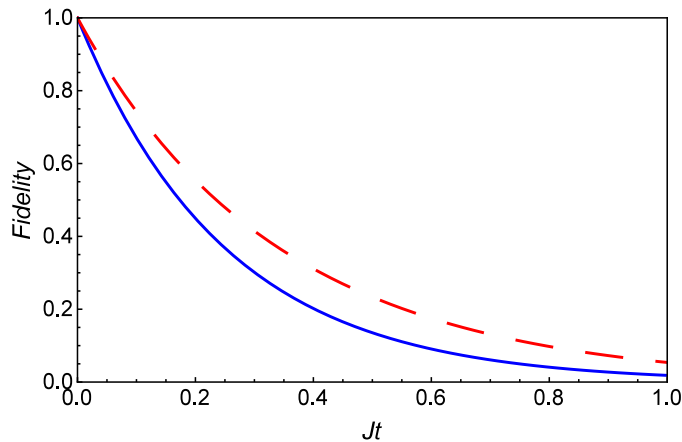}
\caption{The time evolution of Fidelity with respect to the MES in case of
different noise channels. Fidelity for the case of engineered reservoir (red
dashed), given by Eq.(\protect\ref{ME2}), decays slower as compared to the
case of standard reservoir (blue line), described by Eq.(\protect\ref{ME}).
Here $\protect\gamma=J$, $k_a=k_b=J/2$.}
\label{fig4}
\end{figure}

 Under this two photon approximation, the
corresponding Lindblad term of ME can be rewritten as 
\begin{equation}
\mathcal{L}_{\Sigma}=\sum_{k=1}^{3}\frac{\gamma _{k}}{2}(2\Sigma_{k}\rho
\Sigma_{k}^{\dagger }-\{\Sigma_{k}^{\dagger }\Sigma_{k},\rho \}),  \label{ME2}
\end{equation}%
where $\gamma ^{a}=\gamma ^{b}=\gamma $ and we introduced a new collective
operator $\Sigma=ab$. Notice that operator $\Sigma$ includes modes $a$ and $b$ for the
same cavity. Nevertheless, it is easy to see that this approximation for the
interaction with the reservoir it is not equivalent to have the two modes in
a common reservoir, in which case the correct substitution would be $\Sigma=a+b$.
In Fig.\ref{fig4} we compare this new way of modeling the reservoir (\ref%
{ME2}) to the noise channel defined by Eq.(\ref{ME}) by measuring the
Fidelity with respect to the MES. Notice that the MES is more robust against
the decoherence for the \textit{coupled decay} type of reservoir, i.e.
provided that both modes inside each cavity decay cooperatively.

\begin{figure}[t]
\centering
\includegraphics[width=0.5 \textwidth]{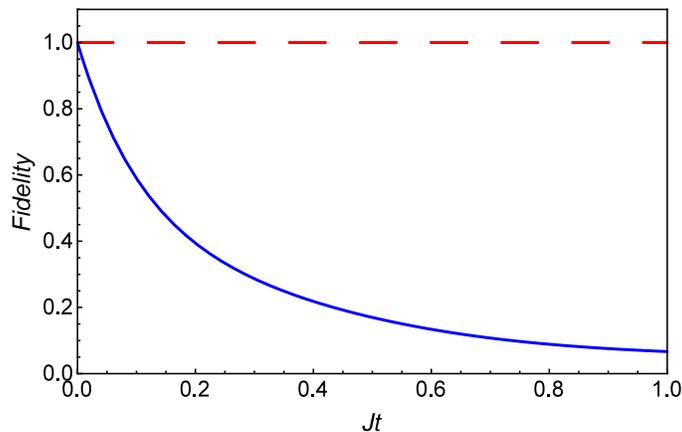}
\caption{As result of the symmetry between the different modes composing the
MES, we observe numerically and analytically, that this state remains
unaltered (red dashed line) under a Phase Flip noise channel that interacts
with the system by Eq.(\protect\ref{ME2}). For an interaction of the type
Eq.(\protect\ref{ME}) the fidelity decays as should be expected. $\protect%
\gamma=J$, $k_a=k_b=J$.}
\label{fig5}
\end{figure}

Nevertheless, dissipation may not be the only source of noise, in fact, a
Phase Flip (PF) mechanism \cite{Nielsen} will lead to a strong decoherence.
It is noticed that the MES state (\ref{MES2}), is very sensitive to
decoherence (losses of the off diagonal elements). Then, the PF mechanism,
which because of its definition leaves the diagonal elements invariant and
only changes the off diagonal elements, could reproduce for example, the
decrease in the Fidelity with respect to the MES state. Going in this
direction, we propose a PF noise channel, by replacing the creation and
annihilation operators in Eq.(\ref{ME}) by,

\begin{equation}
\sigma _{\alpha }=|0\rangle \langle 0|+e^{i\theta }|1\rangle \langle
1|+e^{i\theta }|2\rangle \langle 2|,  \label{phase_p}
\end{equation}%
with $\theta =\pi $. This operator introduces a phase on the state if there
is at least one photon in the mode, otherwise it leaves the state invariant.
It would be interesting now to see how this PF noise acts on the MES by
combining it with the two damping mechanisms defined by Eqs. (\ref{ME}) and (%
\ref{ME2}). This is shown in Fig. \ref{fig5}. For the case of the coupled
decay (\ref{ME2}) we have to replace operator $\Sigma$ by $P=\sigma _{a}\sigma
_{b}$. We observe in Fig. \ref{fig5} that under the coupled decay
approximation for the noise channel (\ref{ME2}), the MES state is completely
robust against the Phase Flip noise channel. On the contrary, for the case
of the noise channel described in (\ref{ME}) the fidelity decays rapidly.


\section{Conclusion}

Entanglement is a fundamental tool for quantum information tasks, however,
and unfortunately, it is very fragile with respect to environmental noise
which poses a problem of not only creating entanglement, but a practical way
of protecting it. In this work, we addressed the problem of generation and
protection of a MES in an optical network. We showed that one can generate a
MES as a zero energy eigenstate of the Hamiltonian of the system, that
satisfies the conditions $(\ref{hopT3})$ and ($\ref{HfT1}$). We also discuss
the dynamics to reach such a state, starting from the ground state of the
system. If we vary the phase in time, with a velocity $\alpha$, it takes a
rapid evolution (non adiabatic) to jump over the avoided crossing gap to
reach the desired state. On the other hand, for a slow time evolution (small 
$\alpha$ ), the system will follow the same original state. In this case, we
observe that when approaching the gap region, the global entanglement
experiences a maximum while the bi and three partite entanglements show a
minimum, providing the evidence that a collective phenomena takes place.
Also, we suggest, via reservoir engineering, how to protect the MES from
decoherence. In particular, we show that for a collective decay model, i.e.
Eq. (\ref{ME2}) , the system is robust under decoherence mechanisms such as
phase flip. Finally, let's discuss the experimental realization of our proposal. We focus first on the hopping term in Eq. (\ref{Hhopp}). The coupling between the cavities can be achieved mainly in two different ways \cite{Orszag2}, either via an optical fiber \cite{Coto} or by tunnel effect \cite{Hartmann}. The hopping we modeled is a general expression that fits well with others model like Bose-Hubbard model \cite{Hartmann} and super conducting qubits \cite{Underwood}, and it can be tuned to be of the order of MHz. We choose to scale all the parameter with the hopping strength ($J$), such that some conclusion of this work can be extended to others system. Then, in order to see the feasibility of the experiment, we are only interested in the rate $k/J$, with $k$ representing the nonlinearities. The Kerr self interaction naturally appears in some media as a result of a non-zero third-order electric susceptibility, but its effect is negligible on the level of few photons. However, the strong interaction of the light mode with atoms inside a cavity QED, under particular circumstances, mediates strong nonlinear interaction among the photons of the cavity mode \cite{Schmidt}. The Kerr cross interaction can be more difficult to realize for photons, but as self interaction it should be done effectively through a nonlinear medium, e.g. , atoms. Furthermore, in the polaritonic basis, both Kerr interactions appears for the case of four level atoms \cite{Hartmann2}, indicating that this approach is possible. Once again, in Bose Hubbard model these nonlinear interactions appears naturally \cite{Reyes}. For a three level atom configuration, the nonlinear interaction can be explained by the Stark shift \cite{Hartmann}, and this shift have been measured experimentally \cite{Hau} to be of the order of MHz. Even more, in recent experiments for super conducting qubits, similar nonlinear strength has been found, for self and cross interaction \cite{Holland}. To conclude, the rates we used in our simulations can be obtained experimentally in a variety of systems. 

\section{Acknowledgment}

The authors would like to thank the support of projects Fondecyt No. 1140994
and Conicyt-PIA anillo ACT-1112 ``Red de analisis estocastico y aplicaciones%
''. One of us (R.C.) also thanks the support of Fondecyt (postdoctoral
fellowship No. 3160154). 


\begin{thebibliography}{99}
\bibitem{Horodecki} R. Horodecki, P. Horodecki, M. Horodecki, K. Horodecki,
Rev. Mod. Phys. \textbf{81}, 865 (2008).

\bibitem{Vedral} V. Vedral, Nat. Phys. \textbf{10}, 256 (2014).

\bibitem{Zhao} M.-J. Zhao, Phys. Rev. A \textbf{91}, 012310 (2015).

\bibitem{Kastoryano} M. J. Kastoryano, F. Reiter and A. S. S\o rensen, Phys.
Rev. Lett. \textbf{106}, 090502 (2011).

\bibitem{Datta} A. Datta, A. Shaji and C.M. Caves, Phys.Rev.Lett \textbf{100}%
, 050502 (2008).

\bibitem{Piani1} M. Piani, P. Horodecki, R. Horodecki, Phys. Rev. Lett. 
\textbf{100}, 090502 (2008).

\bibitem{Xu} J.-S. Xu, \textit{et al.}, Nature Commun. \textbf{1}, 7 (2010);
J.-S. Xu, \textit{et al.}, Nature Commun. \textbf{4}, 2851 (2013).

\bibitem{Piani2} M. Piani, S. Gharibian, G. Adesso, J. Calsamiglia, P.
Horodecki, and A. Winter, Phys. Rev. Lett. \textbf{106}, 220403 (2011).

\bibitem{Streltsov} A. Streltsov, H. Kampermann, and D. Bru\ss , Phys. Rev.
Lett. \textbf{106}, 160401 (2011).

\bibitem{Adesso} G. Adesso, V. D´Ambrosio, E. Nagali, M. Piani, F.
Sciarrino, Phys. Rev. Lett. \textbf{112}, 140501 (2014).

\bibitem{Lanyon} B. P. Lanyon, P. Jurcevic, C. Hempel, M. Gessner, V.
Vedral, R. Blatt, and C. F. Roos, Phys. Rev. Lett. \textbf{111}, 100504
(2013).

\bibitem{Orieux} A. Orieux, M.A. Ciampini, P. Mataloni, D. Bru\ss , M.
Rossi, and C. Macchiavello, Phys. Rev. Lett. \textbf{115}, 160503 (2015).

\bibitem{Nielsen} M. A. Nielsen and I. L. Chuang, \textit{Quantum
Computation and Quantum Information} (Cambridge University Press, Cambridge,
2010).

\bibitem{Zeng} B. Zeng, D. L. Zhou, Z. Xu, C. P. Sun and L. You, Phys. Rev.
A \textbf{71}, 022309 (2005).

\bibitem{Devoret} M. H. Devoret and R. J. Schoelkopf, Science \textbf{339},
1169 (2013).

\bibitem{Reed} M. D. Reed, in Entanglement and Quantum Error Correction with
Superconducting Qubits, Ph.D. thesis, Yale University (2013).

\bibitem{Lidar} D. A. Lidar, I. L. Chuang and K. B. Whaley, Phys. Rev. Lett. 
\textbf{81}, 2594 (1998).

\bibitem{Mundarain} D. Mundarain and M. Orszag, Phys. Rev. A \textbf{75},
040303(R) (2007).

\bibitem{Poyatos} J. F. Poyatos, J. I. Cirac, and P. Zoller, Phys. Rev.
Lett. \textbf{77}, 4728 (1996).

\bibitem{Verstraete} F. Verstraete, M.M. Wolf, and J. I. Cirac, Nat. Phys., 
\textbf{5}, 633 (2009).

\bibitem{Barreiro} J. T. Barreiro, \textit{et al.}, Nature Physics \textbf{6}%
, 943 (2010).

\bibitem{Krauter} H. Krauter, C. A. Muschik, K. Jensen, W. Wasilewski, J. M.
Petersen, J. I. Cirac, and E. S. Polzik, Phys. Rev. Lett. \textbf{107},
080503 (2011).

\bibitem{Murch} K. W. Murch, \textit{et al.}, Phys. Rev. Lett. \textbf{109},
183602 (2012).

\bibitem{Kordas} G. Kordas, S. Wimberger, and D. Witthaut, Europhys. Lett. 
\textbf{100}, 30007 (2012).

\bibitem{Aron} C. Aron, M. Kulkarni, and H. E. T\"{u}reci, Phys. Rev. A 
\textbf{90}, 062305 (2014).

\bibitem{Holland} E. T. Holland, \textit{et al.}, Phys. Rev. Lett. \textbf{%
115}, 180501 (2015).

\bibitem{Leghtas} Z. Leghtas, \textit{et al.}, Science \textbf{347}, 6224
(2015).

\bibitem{Mikhalychev} A. Mikhalychev, D. Mogilevtsev and S. Kilin, J. Phys.
A: Math. Theor. \textbf{44} 325307 (2011).

\bibitem{Mogilevtsev} D. Mogilevtsev, A. Mikhalychev, V. S. Shchesnovich,
and N. Korolkova, Phys. Rev. A \textbf{87}, 063847 (2013).

\bibitem{Chaitanyaa} K.V.S. Shiv Chaitanyaa, S. Ghoshb and V. Srinivasan, J.
Mod. Opt. \textbf{61}, 1409 (2014).

\bibitem{Azouit} R. Azouit, A. Sarlette, P. Rouchon, arXiv:1503.06324 and
arXiv:1511.03898

\bibitem{Heiss} W. D. Heiss and A. L. Sannino, J. Phys. A: Math. Gen. 
\textbf{23}, 1167 (1990).

\bibitem{Bhattacharya} M. Bhattacharya and C. Raman, Phys. Rev. Lett. 
\textbf{97}, 140405 (2006); \textit{ibid}, Phys. Rev. A \textbf{75}, 033405
(2007).

\bibitem{Eleuch} H. Eleuch and I. Rotter, Fortschr. Phys. \textbf{61}, 194
(2013).

\bibitem{Gonzalez} R. Gonzalez-Ferez and J. S. Dehesa, Phys. Rev. Lett. 
\textbf{91}, 113001 (2003).

\bibitem{Karthik} J. Karthik, A. Sharma, and A.Lakshminarayan, Phys. Rev. A 
\textbf{75}, 022304 (2007).

\bibitem{Oh} S. Oh, Z. Huang, U.Peskin, and S. Kais, Phys. Rev. A \textbf{78}%
, 062106 (2008).

\bibitem{Wang} Z. H. Wang, Q. Zheng, Xiaoguang Wang and Yong Li, Sci. Rep. \textbf{6}, 22347 (2016).

\bibitem{Joshi} A. Joshi and S. V. Lawande, Phys. Rev. A \textbf{46}, 5906
(1992).

\bibitem{Harouni} M. J. Faghihi, M. K. Tavassoly and M. B. Harouni, Laser
Phys. \textbf{24}, 045202 (2014).

\bibitem{Yamamoto} N. Imoto, H. A. Haus, Y. Yamamoto, Phys. Rev. A \textbf{32%
}, 2287 (1985).

\bibitem{Aliaga} G. Berlin and J. Aliaga, J. Mod. Opt. \textbf{48}, 1819
(2001).


\bibitem{Landauer} R. Landauer, Phys. Lett. A \textbf{25}, 416 (1967).

\bibitem{Sipe} Y.-K. Yoona, R. S. Benninka, R. W. Boyda and J. E. Sipe, Opt.
Comm. \textbf{179}, 577 (2000).

\bibitem{Reyes} S. A. Reyes, L. Morales-Molina, M. Orszag and D. Spehner, E.
Phys. Lett. \textbf{108}, 20010 (2014).

\bibitem{negativity} G. Vidal and R. F. Werner, Phys. Rev. A \textbf{65},
032314 (2002).

\bibitem{negativity2} T. Baumgratz, M. Cramer and M. B. Plenio, Phys. Rev.
Lett. \textbf{113}, 140401 (2014).

\bibitem{peres} A. Peres, Phys. Rev Lett. \textbf{77}, 1413 (1996).

\bibitem{Cheng} Y-C. Ou and H. Fan, Phys. Rev. A \textbf{75}, 062308 (2007).

\bibitem{Sabin} C. Sabin and G. Garcia-Alcaine, Eur. Phys. J. D \textbf{48},
435 (2008).

\bibitem{Haake} F. Haake, ``\textit{Quantum Signature of Chaos}", Springer,
second edition (2001).

\bibitem{Sachdev} S. Sachdev, ``\textit{Quantum Phase Transitions}",
(Cambridge Univ. Press, 1999).

\bibitem{Wu} L. A. Wu, M. S. Sarandy and D. A. Lidar, Phys. Rev. Lett 
\textbf{93}, 250404 (2004).

\bibitem{Campbell} S. Campbell, L. Mazzola, G. De Chiara, T. J. G. Apollaro,
F. Plastina, Th. Busch and M. Paternostro, New J. Phys. \textbf{15}, 043033
(2013).

\bibitem{Werlang} T. Werlang, C. Trippe, G. A. P. Ribeiro and G. Rigolin,
Phys. Rev. Lett. \textbf{105}, 095702 (2010).

\bibitem{Schumacher} B. Schumacher, Phys. Rev. A \textbf{51}, 2738 (1995).

\bibitem{Orszag} M. Orszag, \textit{Quantum Optics}, Springer, 2nd edition
(2008).

\bibitem{Orszag2} M. Orszag, N. Ciobanu, R. Coto and V. Eremeev, J. Mod. Opt. \textbf{62}, 593 (2015).

\bibitem{Coto} R. Coto and M. Orszag, M. J. Phys. B: At. Mol. Opt. Phys. \textbf{46}, 175503 (2013).

\bibitem{Hartmann} M. J. Hartmann, F. G. S. L. Brandao and M. B. Plenio, Laser \& Photon Rev. \textbf{2}, 527 (2008).

\bibitem{Underwood} D. Underwood, W. Shanks, J. Koch, and A. A. Houck, Phys. Rev. A \textbf{86}, 023837 (2012).	

\bibitem{Schmidt} H. Schmidt and A. Imamoglu, Opt. Lett. \textbf{21}, 1936 (1996).

\bibitem{Hartmann2} M. J. Hartmann, F. G. S. L. Brandao and M. B. Plenio, Nature Phys. \textbf{2}, 849 (2006).

\bibitem{Hau}L. V. Hau, S. E. Harris, Z. Dutton, and C. H. Behroozi, Nature (London) \textbf{397}, 594 (1999).




\end{thebibliography}
\end{document}